\documentclass[letterpaper,11pt]{article}
\usepackage[top=2cm, bottom=2.5cm, left=2.5cm, right=2.5cm]{geometry}

\usepackage{amssymb,amsthm,graphicx,amsfonts,amssymb,amsmath,float,subcaption,cite}
\usepackage[pdftex]{hyperref}

\usepackage{xcolor}
\usepackage{chngcntr}

\usepackage{stackengine}
\stackMath
\usepackage{xcolor}

\newcommand{\norm}[1]{\left\lVert#1\right\rVert}

\definecolor{cinvestav}{RGB}{0, 159, 147}

\title{Freezable bound states in the continuum for time-dependent quantum potentials}

\author{Izamar Guti\'errez Altamirano$^1$, Alonso Contreras-Astorga$^2$, Alfredo Raya$^{1,3}$ \\
\vspace{1mm}
\\
	$^1$\textit{Instituto de F\'isica y Matem\'aticas, Universidad Michoacana de San Nicol\'as} \\ 
	\textit{de Hidalgo, Edificio C-3, Ciudad Universitaria.  
	Francisco  J. M\'ujica s/n. }\\ 
	\textit{Col. Fel\'icitas del R\'io. 58040 Morelia, Michoac\'an, M\'exico.} \\
	$^2$\textit{CONACyT - Physics Department, Cinvestav, P.O. Box. 14-740,}\\
	\textit{07000 Mexico City, Mexico.}\\
    $^3$\textit{Centro de Ciencias Exactas, Universidad del B\'{\i}o-B\'{\i}o,  Avda. Andr\'es Bello 720,} \\ 
    \textit{Casilla 447, 3800708, Chill\'an, Chile.}
    \\
	\\\sl{\small{E-mails: izamar.gutierrez@umich.mx, alonso.contreras@conacyt.mx, alfredo.raya@umich.mx}} }

\date{}

\begin{document}
\maketitle
% \begin{frontmatter}

% \title{Freezable bound states in the continuum for quantum time dependent potentials}

% \author[1]{Izamar Guti\'errez Altamirano}
% \ead{izamar.gutierrez@umich.mx}

% \author[2]{Alonso Contreras-Astorga\corref{cor1}}
% \ead{alonso.contreras@conacyt.mx}

% \author[1,3]{Alfredo Raya}
% \ead{alfredo.raya@umich.mx}
% \address[1]{Instituto de F\'isica y Matem\'aticas, Universidad Michoacana de San Nicol\'as de Hidalgo, Edificio C-3, Ciudad Universitaria. Francisco J. M\'ujica s/n. Col. Fel\'icitas del R\'io. 58040 Morelia, Michoac\'an, M\'exico}
% \address[2]{CONACyT - Physics Department, Cinvestav, P.O. Box. 14-740, 07000 Mexico City, Mexico}
% \address[3]{Centro de Ciencias Exactas, Universidad del B\'{\i}o-B\'{\i}o, Avda. Andr\'es Bello 720, Casilla 447, 3800708, Chill\'an, Chile}

% \cortext[cor1]{Corresponding author}

\begin{abstract}

In this work, we construct time-dependent potentials for the Schr\"odinger equation via supersymmetric quantum mechanics. The generated potentials have a quantum state with the property that after a particular threshold time $t_F$, when the potential does no longer change, the evolving state becomes a bound state in the continuum, its probability distribution freezes. After the factorization of a geometric phase, the state satisfies a stationary Schr\"odinger equation with time-independent potential. The procedure can be extended to support more than one bound state in the continuum. Closed expressions for the potential, the bound states in the continuum, and scattering states are given for  the examples starting from the free particle.  

\vspace{1mm}

\end{abstract}

% \begin{keyword}
% %% keywords here, in the form: keyword \sep keyword
%  Bound states in the continuum \sep Supersymmetric quantum mechanics 

% %% PACS codes here, in the form: \PACS code \sep code

% %% MSC codes here, in the form: \MSC code \sep code
% %% or \MSC[2008] code \sep code (2000 is the default)

% \end{keyword}

% \end{frontmatter}

%% \linenumbers

%% main text
\section{Introduction}\label{intro}

Bound states in the continuum (BICs) were first discussed in quantum mechanics in the seminal work of von Neumann and Wigner~\cite{1929PhyZ...30..467V}, where they construct a localized, normalizable zero-mode state of the form $\psi(r)=\sin(r^3)/r^2$ in the potential $V(r)=2r^{-2}+9r^{4}$, which admits only continuum spectrum solutions for non-vanishing energy eigenvalues. These authors further considered a wave function including modulation of the free particle profile.  By analyzing the behavior of the mode as $r\to \infty$ with an energy $E\ge 0$ embedded in the continuum, they construct a periodic potential from the modulated wave function $V(r)=E+\nabla^2\psi(r)/\psi(r)$ such that $V(r\to\infty)\sim-|V_0|<E$. Upon demanding the normalizability of the state, the potential hence constructed exhibits oscillatory behavior with half the period of the wave function, in such a way that the localization required for the normalization of the state can be understood as the result of its reflection on the {\em Bragg mirror} generated by the crests of the oscillation of the potential. The family of von Neumann-Wigner potentials has been continuously revisited for almost a century (see, for instance, Refs.~\cite{SimonCPAM69,PhysRevA.11.446,GAZDY197789}). It is known that potentials of the form $V(r)=a \sin(br)/r$ admit a BIC at energy $E=b^2/4$ provided $|a|>|b|$~\cite{klaus91}. These quantum states have been studied under several approaches, including the  Gelfan-Levitan equation~\cite{GelfanLev51-2} or inverse scattering approach~\cite{GAZDY197789,PhysRevA.50.4478}, Darboux transformations~\cite{PhysRevA.51.934,LohrRMF18,Nicolas20}, supersymmetry (SUSY)~\cite{PhysRevA.48.3525,DEMIC20152707,Nicolas13}, among others. Nevertheless, nowadays, BICs are recognized as a general wave phenomenon~\cite{Hsu16} explored in various scenarios, including atomic physics~\cite{PhysRevA.11.446,PhysRevA.10.1122,PhysRevA.31.3964}, optic waveguides~\cite{Longhi:14}, acoustics~\cite{ParkerJSV66,lyapina_maksimov_pilipchuk_sadreev_2015}, and even water waves~\cite{LintowWM07}.  Special interest deserves the development of such states in materials, ranging from photonics to quantum dots in a plethora of different setups and aiming for technological applications (see~\cite{Hsu16} for a review). BICs have also been studied in graphene~\cite{Gonz_lez_2010}, some topologically insulating materials~\cite{SablikovPLA379} and, from the formal point of view, modeling the Dirac equation in curved space~\cite{Gosh21}. A common denominator in these cases is the static character of the effective potential in the effective wave equation governing the underlying system.

Although the completeness of the continuum spectrum of a wave operator might suggest that in principle, any localized squared integrable wave function can be expressed as a combination of these states, in the case of the Schr\"odinger equation, one has to be careful as far as the realization of BICs is concerned. For instance, in the case of time-dependent potentials, when the time evolution of the potential is frozen, it is not guaranteed that combinations of this kind are automatically solutions to the stationary Schr\"odinger equation, as we present in this article.
The time-dependent Schr\"odinger equation can be solved exactly in a handful of cases, such as potential wells with moving walls~\cite{LawrenceWheller53,DoescherAJP37,Contreras-Astorga19}. Several approximations are known to explore the analytical properties of the time-dependent wavefunctions and energy eigenvalues (see, for instance, Ref.~\cite{cooney2017infinite} and references therein), including the adiabatical approximation~\cite{Berry84} and perturbation theory~\cite{DoescherAJP37}. A powerful strategy to construct time-dependent solutions to the Schr\"odinger equation from its stationary version is through point transformations~\cite{PhysRevA.26.729,BlumanSIAM43,Zelaya17,Zelaya19,Cruz20}. These transformations, in combination with first-order time-independent SUSY, have allowed extending the number of solvable time-dependent examples, from the infinite potential well with a moving wall to the trigonometric P\"oschl-Teller potential~\cite{Contreras-Astorga19}, by transforming the stationary Schr\"odinger equation to a time-dependent equation that in the remote past/future connects with the solutions of the free particle.

This article presents a general framework for constructing time-dependent potentials for the Schr\"odinger equation employing second-order supersymmetry in combination with point transformations. We build a BIC for the time-dependent case by point-transforming the stationary problem, modifying the potential and wavefunction. Then, we assume that after a specific time, all the time dependence of the potential is frozen, such that the potential becomes once more stationary and, we explore the behavior of the normalizable state. Intriguingly, it is seen that the freezable BIC is not an eigensolution of the stationary Schr\"odinger equation in the frozen potential but rather solves an equation that includes a vector potential that does not generate a magnetic field, nevertheless. Thus, by an appropriate gauge transformation, we gauge away the vector potential and observe the BIC that remains frozen when the potential ceases to evolve in time. We exemplify these features starting with the wave function of a free particle in the real positive semi-axis. We further extend this system by constructing a second BIC hence illustrating the procedure to find more intricate systems that  support a finite~\cite{BSimonAMS97,NabokoTMP86} and infinite number~\cite{GAZDY197789,MVerAJP82,Pivovarchik_1986} of BICs. To this end, we have organized the remaining of this article as follows: In Section~\ref{Preliminaries} we describe the preliminaries of SUSY, point transformations, gauge invariance, and geometric phases in our framework. Section~\ref{Frezable} presents the setup for BICs and how to freeze them within the framework. Explicit examples are discussed in Section~\ref{Examples}, and final remarks are presented in Section \ref{FR}.

%%%%%%%%%%%%%%%%%%%%
\section{Preliminaries}\label{Preliminaries}
%%%%%%%%%%%%%%%%%%%%

Before going into the general technique, let us introduce the three main tools we need to generate time-dependent potentials with freezable bound states in the continuum. First, confluent supersymmetric quantum mechanics allows modifying the spectrum of a quantum Hamiltonian. A point transformation provides dynamics into the system. Finally, a gauge transformation facilitates the interpretation of the results.

\subsection{Supersymmetric Quantum Mechanics}
Supersymmetric Quantum Mechanics or SUSY is a technique that allows us to find solutions of a Schr\"odinger equation given that we know a solution of another Schr\"odinger equation with different potential term \cite{matveev92,Khare95,Fernandez04,Andrianov04,gangopadhyaya17,junker19}. 
In it simplest form, we consider a one dimensional quantum Hamiltonian $H_0$ and a first-order differential operator $L_1^\dag$ that maps solutions of the eigenvalue equation $H_0 \psi~=~E \psi$ into solutions of $H_1 \hat{\psi}=E \hat{\psi}$, where $H_1$ is a Hamiltonian with a deformed potential term. In this work, we consider confluent supersymmetry, which is a second-order SUSY that can be seen as two iterations of first-order transformations \cite{Baye1993,Sparenberg1995,PhysRevA.51.934,Boya1997,Mielnik2000,Rosu2000,C2003,FernandezC.2005,C2011,Bermudez2012,Schulze-Halberg2013,Contreras15,Contreras17}. For sake of completeness, let us review the necessary parts of the formalism required for this work. We start out by considering a Hermitian Hamiltonian $H_0$ with a time-independent potential $V_0(y)$ that could have discrete, continuous  spectra, or both. Also, we consider as known some solutions of the eigenvalue equation: 
\begin{eqnarray}
H_0 \psi = E \psi,  \quad\text{where} \quad H_0= - \frac{d^2}{dy^2} + V_0(y), 
\end{eqnarray}
$y \in (y_\ell,y_r)\subset \mathbb{R}$ and $E$ is the real energy parameter. Then, we apply two very specific steps of 1-SUSY to arrive to a confluen SUSY transformation.  

\subsubsection{First-order supersymmetry}

As a first step, we propose the intertwining relation
\begin{eqnarray}
H_1 L_1^\dag = L_1^\dag H_0,  \label{intertwining}
\end{eqnarray}
where 
\begin{eqnarray}
H_1= - \frac{d^2}{dy^2} + V_1(y), \qquad L_1^\dag = - \frac{d}{dy} + \frac{u'}{u}, \label{def H1}
\end{eqnarray}
$u=u(y)$ is a function to be found called \emph{seed} or \emph{transformation function}. By substituting \eqref{def H1} into the intertwining relation \eqref{intertwining} we find that $V_1$ and $u$ must fulfill 
\begin{eqnarray} \label{potential V1}
V_1(y)= V_0(y) - 2 \frac{d^2}{dy^2} \ln u, \quad  -u''+V_0 u = \epsilon u, 
\end{eqnarray}
where $\epsilon$ is a real integration constant called \emph{factorization energy}. Notice that $u$ satisfies the initial Schr\"odinger equation for the energy parameter $\epsilon$ but we do not impose on it the boundary conditions of the initial physical problem. From expression \eqref{potential V1} we see that to have a regular potential $V_1$ the transformation function  $u$ must not vanish.  By applying the operator $L_1^\dag$ onto a solution $\psi$ we can obtain eigenfunctions $\hat{\psi} \propto L_1^\dag \psi$ of the Hamiltonian $H_1$, the intertwining equation \eqref{intertwining} guarantees this. In other words, the operator $L_1^\dag$ maps the space of solutions of $H_0 \psi = E \psi$ onto the space of solutions $H_1 \hat{\psi} = E \hat{\psi}$.  An inverse map exists and can be found from the formally adjoint intertwining relation~\eqref{intertwining}, $H_0 L_1 = L_1 H_1$, where $L_1 = (L_1^\dag)^\dag= d/dy+u'/u$. Note that operators $L_1$ and $L_1^\dag$ factorize the Hamiltonians $H_0$ and $H_1$ as $L_1 L_1^\dag = H_0 - \epsilon$,  $L_1^\dag L_1 = H_1 - \epsilon$. Such factorizations are useful to find the \emph{missing state} of $H_1$, it is the state annihilated by $L_1$, and it satisfies the equation $H_1 \hat{\psi}_\epsilon = \epsilon \hat{\psi}_\epsilon$. By solving the first-order differential equation $L_1 \hat{\psi}_\epsilon=0$ we find that 
\begin{eqnarray} \label{missing 1susy}
\hat{\psi}_\epsilon(y) = \hat{c}_\epsilon \frac{1}{u(y)},
\end{eqnarray}
where $\hat{c}_\epsilon$ is a normalization constant when $\hat{\psi}_\epsilon$ is normalizable, otherwise $\hat{c}_\epsilon = 1$. 
Now,  any other solution of $H_1 \hat{\psi}=E \hat{\psi}$ besides $\hat{\psi}_\epsilon$ can be obtained as $\hat{\psi} \propto L_1^\dag \psi$. Because we know that $L_1 L_1^\dag = H_0 - \epsilon$, we can calculate the normalization constant. Assuming $\norm{\psi}^2=1$, we see that 
\begin{eqnarray} \label{eigenfunctions 1susy}
\hat{\psi} = \frac{1}{\sqrt{E - \epsilon}} L_1^\dag \psi.
\end{eqnarray}

Thus, starting from a Schr\"odinger equation $H_0 \psi = E \psi$ with potential $V_0$, and solutions $\psi$, we obtain a Schrodinger equation $H_1\hat{\psi} = E \hat{\psi}$ with a potential $V_1$, and solutions $\hat{\psi}_\epsilon$ and $\hat{\psi}$, see  \eqref{potential V1}, \eqref{missing 1susy} and  \eqref{eigenfunctions 1susy}. 

%%%%%%%%%%%%%%%%%%%
\subsubsection{Second-order confluent supersymmetry} \label{Sec Confluent}
%%%%%%%%%%%%%%%%%%%

We can iterate the procedure and obtain a second Hamiltonian $H_2$. The selection of the transformation function $v$ and the factorization energy $\varepsilon$ by properly  solving $H_1 v = \varepsilon v$, will fix $H_2$ and consequently its eigenfunctions. There exist many variations of the second iteration leading to different systems. The most common is when $\epsilon \neq \varepsilon$ are taken both as real constants. Here we consider the case $\epsilon = \varepsilon \in \mathbb{R}$ . Once we fixed the factorization constant, we need to select a transformation function $v$. %solving $H_1 v = \epsilon v$. 
A reasonable choice is the missing state $v =  1/ u$, but this choice results in $H_2 = H_0$. Since we want to generate a Hamiltonian different from the initial one, $H_2 \neq H_0$, we need to use a general solution of $H_1 v = \epsilon v$; this can be done using the reduction of order formula,
\begin{eqnarray}
v=\frac{1}{u}\left(\omega +  \int^y_{y_0} u^2(z) dz     \right),
\end{eqnarray} 
where $\omega$ is a real constant. The potential associated to this second iteration becomes
\begin{eqnarray} \label{SUSY C V2}
V_2(y) = V_1(y) - 2 \frac{d^2}{dy^2}\ln v = V_0(y) - 2 \frac{d^2}{dy^2} \ln  \left(\omega +  \int^y_{y_0} u^2 dz   \right).
\end{eqnarray}
The Hamiltonians $H_0$ and $H_2$ are intertwined by the operator 
\begin{eqnarray} \label{SUSY L}
L^\dag = L_2^\dag L_1^\dag = \left(- \frac{d}{dy}+ \frac{v'}{v}  \right) \left(- \frac{d}{dy} + \frac{u'}{u}   \right),
\end{eqnarray}
as $H_2 L^\dag = L^\dag H_0$. 

Solutions of the equation $H_2 \bar{\psi} = E \bar{\psi}$ can be found applying $L^\dag$ onto solutions of $H_0 \psi = E \psi$ as 
\begin{eqnarray} \label{bar psi}
\bar{\psi} = \frac{1}{E - \epsilon} L^\dag \psi.  
\end{eqnarray}
Moreover, the associated missing state to the factorization energy $\epsilon$ is 
\begin{eqnarray} \label{bar missing}
\bar{\psi}_\epsilon = C_\epsilon \frac{1}{v} = C_\epsilon \frac{u}{\omega +  \int^y_{y_0} u^2 dz },
\end{eqnarray}
where $C_\epsilon$ is a normalization constant, if $\bar{\psi}_\epsilon$ is square integrable. In the first-order SUSY, the transformation function $u$ must be nodeless to produce a regular potential $V_1$. In the confluent case this restriction changes, the function $\omega + \int_{y_0}^y u^2 dz$ must not vanish. We can fulfill this requirement selecting $u$ such that either $\lim_{y \to y_\ell} u(y) =0$ or $\lim_{y \to  y_r} u(y) =0$, then we can guarantee that there exist constants $\omega$ and $y_0$ that keep $V_2$ regular.

%%%%%%%%%%%%%%%%%%%%%%%%%%%%
%%%%%%%%%%%%%%%%%%%%%%%%%%
\subsection{Point transformation} \label{Point T}
%%%%%%%%%%%%%%%%%%%%%%%%%%%%
%%%%%%%%%%%%%%%%%%%%%%%%%%%%
We can relate a one-dimensional time-independent Schr\"odinger equation with a time-dependent one using a point transformation, see for example \cite{PhysRevA.26.729,BlumanSIAM43}. Moreover, the connection between a time-dependent Supersymmetry presented in \cite{matveev92,Bagrov95} and the time-independent version was done in \cite{Finkel99}. The combination of Supersymmetric QM and point transformations was further exploited in      \cite{Jana08,Suzko09,Schulze09,Schulze14,Contreras-Astorga19}. In particular, let us consider the time-independent Schr\"odinger equation of the SUSY partner potential $V_2(y)$:
\begin{eqnarray}
\frac{d^2}{dy^2}\bar{\psi}(y) + (E- V_2(y))\bar{\psi}(y)=0.  \label{Finkel TISE}
\end{eqnarray}
Now, let us consider arbitrary functions $A=A(t)$ and $B=B(t)$ and let the variable $y$ be defined in terms of a temporal parameter $t$ and a new spatial variable $x$ as:
\begin{eqnarray}
y(x,t)= x \exp\left[ 4 \int_{t_0}^t A(\tau) d\tau   \right] + 2 \int_{t_0}^t B(\tau) \exp\left[ 4 \int_{\tau_0}^{\tau} A(\tilde{\tau}) d\tilde{\tau}   \right] d\tau, \label{FinkelVariable} 
\end{eqnarray}
then the function
\begin{eqnarray}
\phi(x,t)&=& \hat{\psi}(y(x,t)) \exp \left\{-i \left[ A(t)x^2 + B(t)x + E \int_{t_0}^t \exp\left[8 \int_{\tau_0}^{\tau} A(\tilde{\tau}) d\tilde{\tau} \right] d\tau     \right.  \right. \nonumber \\
& & \left. \left.  + \int_{t_0}^t \left[2 i A(\tau)+B^2(\tau)   \right]d\tau  \right]    \right\}, \label{FinkelFunction}
 \end{eqnarray}
%where $C=C(t)$ is another arbitrary function, 
is solution of the equation
\begin{eqnarray}\label{Finkel TDSE}
i \frac{\partial}{\partial t} \phi(x,t) + \frac{\partial^2}{\partial x^2} \phi(x,t) - V(x,t) \phi(x,t)=0.
\end{eqnarray} 
The former is a time-dependent Schr\"odinger equation (TDSE) where the potential is given by 
\begin{eqnarray}
V(x,t)&=& \exp \left[8 \int_{t_0}^t A(\tau)d\tau  \right] V_2(y(x,t)) + \left[ \frac{d}{dt}A(t)-4 A^2(t) \right] x^2 \nonumber \\
& & + \left[ \frac{d}{dt}B(t) - 4 A(t) B(t)  \right] x. % + C(t). 
\label{FinkelPotential}
\end{eqnarray}
In the last expression we have three terms. The first one involves the potential $V_2$ in terms of $x,~t$ with a time-dependent coefficient. The last two terms are a quadratic and linear monomials in the $x$ coordinate with time-dependent coefficients. Let us set those two last terms equal to zero with the goal to obtain a potential $V$ with a shape similar to the potential $V_2$. This problem was studied in \cite{Contreras-Astorga19}. Setting such coefficients to zero we obtain a system of equations; its solution by direct integration is:  
% \begin{align} \label{ABC}
% &\frac{d}{dt}A(t)-4 A^2(t)=0, &\quad \Rightarrow \quad & A(t)=- \frac{1}{4t +c_1}; \nonumber \\
% & \frac{d}{dt}B(t) - 4 A(t) B(t) = 0, &\quad \Rightarrow \quad & B(t)=  \frac{c_2}{4t + c_1};%\\
% %&C(t)=0;  
% \end{align}
\begin{align} \label{ABC}
A(t)=- \frac{1}{4t +c_1}, \qquad   B(t)=  \frac{c_2}{4t + c_1},
\end{align}
where $c_1$ and $c_2$ are real constants. Once these two functions are known, the change of variable defined in \eqref{FinkelVariable} can be evaluated, 
\begin{eqnarray}
y(x,t)= \frac{2 x - c_2}{2(4t + c_1)}. \label{yvalue}
\end{eqnarray}
Thus, given the stationary Schr\"odinger equation $\bar{\psi}'' + (E-V_2(y))\bar{\psi} = 0$ %where the  potential $V_2=V_2(y)$,
we can find a solution of the time dependent Schr\"odinger equation $i \partial_t \phi + \partial_{xx} \phi - V \phi =0$ where
\begin{align}
&\phi(x,t) = \frac{1}{\sqrt{4t+c_1}} ~\bar{\psi}\left( \frac{2x + c_2}{2(4t+c_1)} \right) \exp \left\{ \frac{i}{4t +c_1}\left[ x^2 - c_2 x + \frac{E+c_2^2}{4}  \right]   \right\}, \label{point phi} 
\end{align}
and
\begin{align}
&V(x,t)= \frac{1}{(4t+c_1)^2} V_2\left( \frac{2x + c_2}{2(4t+c_1)} \right). \label{point V}
\end{align}
There is a singularity we must avoid at $t=-c_1/4$, thus the time domain cannot be whole real line. The domain of the $x$ coordinate could be the same of the $y$  variable, depending on the physical system.

%%%%%%%%%%%%%%%%%%%%%
%%%%%%%%%%%%%%%%%%%%%
\subsection{Gauge invariance and geometric phase} \label{Sec Gauge}
%%%%%%%%%%%%%%%%%%%%
%%%%%%%%%%%%%%%%%%%%

The Schr\"odinger equation for a particle with charge $q$ in an electromagnetic potential is written in terms of the scalar $\varphi$ and vector potential $\mathbf{A}$ rather than in terms of the electric $\mathbf{E}$ and magnetic $\mathbf{B}$ fields by writing the Hamiltonian as $H=(\hat{\mathbf{p}}+q\mathbf{A})^2+q\varphi$.
% \begin{equation}
%     H=(\hat{\mathbf{p}}+q\mathbf{A})^2+q\varphi.
% \end{equation}
Gauge invariance of Maxwell equations implies that the electric and magnetic fields
\begin{equation}
    \mathbf{E}=-\nabla\varphi-\frac{\partial\mathbf{A}}{\partial t}, \qquad \mathbf{B}=\nabla\times \mathbf{A}
\end{equation}
do not change if the following transformations are performed simultaneously,
\begin{equation}
    \mathbf{A}\to \mathbf{A}'=\mathbf{A}+\nabla\lambda, \qquad \varphi\to\varphi'=\varphi-\frac{\partial \lambda}{\partial t},\label{gt}
\end{equation}
where $\lambda=\lambda(x,t)$ is a scalar function. The time-dependent Schr\"odinger equation $ i \partial_t \psi= H \psi$ retains this feature if besides the transformations in Eq.~(\ref{gt}), the wave function changes according to $    \psi\to \psi'=e^{i\lambda}\psi$.
This allows selecting $\lambda$ in such a way that if at a certain instant of time $t_F$ the vector potential $\mathbf{A}\ne 0$ but before we had $\mathbf{A}=0$, one can still have a Schr\"odinger equation without vector potential by tuning the scalar potential appropriately. In particular, by selecting $ \lambda(x,t)=g(x)\Theta(t-t_F)$, 
we can shift the scalar potential in such a way that the time-dependent equation governing this state never develops a vector potential.

Furthermore, in the situation where the time-dependent Schr\"odinger equation
\begin{equation}
    i\frac{\partial \psi'}{\partial t} = (\hat{\mathbf{p}}+q\mathbf{A})^2\psi'+q\varphi \psi'
\end{equation}
involves a vector potential $\mathbf{A}$ such that $\nabla\times \mathbf{A}=0$, we can directly factorize a {\em geometric phase} $ \psi'=e^{i g}\psi$ with
\begin{equation}
    g=q \int \mathbf{A}\cdot d\mathbf{x},
\end{equation}
where $g$ does not depend on the path of integration in the region where the curl of $\mathbf{A}$ vanishes, in such a way that the function $\psi$ verifies
\begin{equation}
    i\frac{\partial \psi}{\partial t} = \hat{\mathbf{p}}^2\psi+q\varphi \psi,
\end{equation}
namely, a time-dependent Schr\"odinger equation without vector potential.

%%%%%%%%%%%%%%%%
%%%%%%%%%%%%%%%%
\section{Freezable bound states in the continuum}\label{Frezable}
%%%%%%%%%%%%%%%%%%%%%
%%%%%%%%%%%%%%%%%%%%

Our goal is to construct a solvable time dependent potential. This potential will change in time until a stopping or freezing time $t_F$, then it will no longer vary in time: 
 \begin{eqnarray} \label{Sec 3 VF}
    V_F(x,t) =  \begin{cases} 
      V(x,t) & 0 \leq t< t_F, \\
      V(x,t_F) &  t\geq t_F. 
  \end{cases}
    \end{eqnarray}
We ask this potential to have at least one BIC when  $t\geq t_F$, we call these states \emph{freezable bound states in the continuum}. 

We start out from a solvable and stationary potential $V_0(y)$ with continuum spectrum. Then, the first step is to construct its confluent SUSY partner $V_2(y)$ using the algorithm presented in Section \ref{Sec Confluent}. The factorization energy $\epsilon$ must be in the continuum spectrum of $H_0$. As a consequence,  the seed solution is a non-normalizable function. Let us study first the case when the domain of the potential is the whole real line. For this case $V_0(y)$ must be a bounded potential. Moreover, we ask 
\begin{eqnarray}
\lim_{y \rightarrow - \infty} V_0(y) = V_\ell, \quad \lim_{y \rightarrow  \infty} V_0(y) = V_r, \qquad V_\ell \neq V_r. 
\end{eqnarray}
Without loss of generality, we will consider $V_\ell > V_r$.  Other requirements that we impose to $V_0(y)$ are:
\begin{align}
\int_{y_a} ^\infty |V_0'|^2 dy < \infty,  \quad \int_{y_a}^\infty |V_0''| dy < \infty, \quad \int^{y_b}_{-\infty} |V_0'|^2 dy < \infty, \quad \int^{y_b}_{-\infty} |V_0''| dy < \infty, \label{inequalities}
\end{align}
where $y_a,~y_b$ are constants with absolute value  arbitrarily large.

Since the asymptotic behaviour of the solutions of the Schr\"odinger equation are important, let us review some general results that will be used. It is known (see, e.g. Theorem 4.6, p.84 in \cite{Berezin91}), that the solutions of the equation $-\psi'' + v_0\psi=k^2 \psi$, for a function $v_0(y)$ satisfying $v_0 \rightarrow 0$ as $y \rightarrow \infty$ and 
\begin{eqnarray}
\int_{y_a} ^\infty |v_0'|^2 dy < \infty, \quad \int_{y_a}^\infty |v_0''| dy < \infty, \label{inequalities 2}
\end{eqnarray}
have the following asymptotic form as $y\rightarrow \infty$:
 \begin{align} \label{theorem asymtotics a}
 &\psi^+(y)= \exp\left( i k \int_{y_a}^y \sqrt{1-\frac{v_0(z)}{k^2}} dz \right) \left(1+ o(1) \right), \nonumber \\ 
 &\phi^+(y)= \exp\left( - i k \int_{y_a}^y \sqrt{1-\frac{v_0(z)}{k^2}} dz \right) \left(1+ o(1) \right).
 \end{align}
Since the substitution of $y$ by $-y$ changes neither the form of the conditions \eqref{inequalities 2} nor the equation, then the solution behaves asymptotically when $y\rightarrow -\infty$ as  
\begin{align}\label{theorem asymtotics b}
&\psi^-(y)= \exp\left( i k \int^{y_b}_{y} \sqrt{1-\frac{v_0(z)}{k^2}} dz \right) \left(1+ o(1) \right), \nonumber \\ 
&\phi^-(y)= \exp\left( -i k \int^{y_b}_y \sqrt{1-\frac{v_0(z)}{k^2}} dz \right) \left(1+ o(1) \right).
\end{align}

The next consideration is to select a factorization energy $V_\ell > \epsilon > V_r$. To apply the results \eqref{theorem asymtotics a} we identify $v_0 =V_0-V_r$ when analysing the asymptotic at $y\rightarrow \infty$, then 
\begin{align} \label{oscillatory}
&\psi^+(y) \propto  \exp \left(i \sqrt{\epsilon - V_r} ~ y \right)\left(1+ o(1) \right), \quad \phi^+(y) \propto \exp\left( - i \sqrt{\epsilon- V_r} ~y \right) \left(1+ o(1) \right), 
\end{align}
i.e., we will have only oscillatory solutions. We must select a seed function $u$ as a superposition, such that $u$ is a real function. We can write $u\propto \sin \left( \sqrt{V_\ell - \epsilon} ~ y + \delta  \right)$, where $\delta$ is a phase. When studying the asymptotic behavior of $u$ at $y\rightarrow -\infty$, we can use $v_0 =V_0-V_\ell$. From  \eqref{theorem asymtotics b}, the solutions of the Schr\"odinger equation have the form
\begin{align} 
&\psi^-(y) \propto  \exp \left( \sqrt{V_\ell - \epsilon } ~ y \right)\left(1+ o(1) \right), \quad \phi^-(y) \propto \exp\left( - \sqrt{V_\ell - \epsilon}~ y \right) \left(1+ o(1) \right), 
\end{align}
the possible behaviour of $u$ is a superposition of a divergent and a convergent exponential functions. %, $\psi^-(y)$ and $\phi^-(y)$. 
We must choose only the convergent solution, $u \propto \psi^-(y)$. By choosing $u$ with these behaviours as $|y| \rightarrow \infty$, we guarantee that there are ranges for the parameters $y_0$ and $\omega$ in \eqref{SUSY C V2} where $V_2$ is a regular potential. 

Moreover, when $y \rightarrow \infty$ then 
\begin{eqnarray}
V_2 \sim V_0  +2 \left\{\frac{2 k  \sin (2 (k y +\delta ))}{\frac{\sin (2 (k y +\delta ))}{2 k }+y+\omega_a}+\frac{\left[\cos (2 (k y +\delta ))+1\right]^2}{\left[\frac{\sin (2 (k y +\delta ))}{2 k }+y+\omega_a\right]^2}\right\} \rightarrow V_r,
\end{eqnarray}
where $k^2=\epsilon - V_r$, and $\omega_a$ is a constant. The missing state behaves as
\begin{eqnarray}
\bar{\psi}_\epsilon \sim \frac{\sin(k y + \delta)}{\frac{\sin (2 (k y +\delta ))}{2 k }+y+\omega_1} \rightarrow 0.  
\end{eqnarray}
When $y \rightarrow -\infty$ the potential and the missing states behave as  
\begin{eqnarray}
V_2 \sim V_0 -\frac{16 \kappa ^3 \omega_b e^{2 \kappa y}}{\left(e^{2 \kappa y}+2 \kappa \omega_b\right)^2} \rightarrow V_\ell, \quad \text{and} \quad \bar{\psi}_\epsilon \sim \frac{e^{\kappa y}}{\frac{e^{2 \kappa y}}{2 k}+\omega_b} \rightarrow 0, 
\end{eqnarray}
where $\kappa^2=V_\ell -\epsilon$. Thus, the potentials $V_0$ and $V_2$ have the same limit as $y \rightarrow \pm \infty$. Moreover, these results suggest that the function $\bar{\psi}_\epsilon$ is square integrable solution of $H_2 \bar{\psi}_\epsilon = \epsilon \bar{\psi}_\epsilon$ with an eigenvalue embedded in the continuum spectrum, in other words, it could be a BIC. Such situations have being discussed in \cite{PhysRevA.48.3525}.

If we start from a potential $V_0(y)$ defined in the semiaxis $(0,\infty)$, the potential can be either bounded or unbounded, but we still require $V_0(y) \rightarrow V_r$ as $y \rightarrow \infty$. Moreover, $V_0(y)$ must satisfy the first two conditions in \eqref{inequalities}.  Then, the correct choice of factorization energy is $\epsilon > V_r$, as part of the continuum spectrum. The behavior of $u$ when $y \rightarrow \infty$ is oscillatory, as explained in \eqref{oscillatory}. Again, it is necessary to choose a real solution. The behavior of $u$ on the left must be chosen so $u$ does not diverge. In fact, $\lim_{y\rightarrow 0} u(y)=0$ is needed, so the missing state $\bar{\psi}_\epsilon$ could satisfy the physical boundary conditions $\bar{\psi}_\epsilon (0)=0$.  

Now, we can associate to $V_2(y)$ a time-dependent potential $V(x,t)$ via the point transformation defined in Section \ref{Point T}. In equation \eqref{point phi} we can see how any solution of $H_2 \bar{\psi}=E \bar{\psi}$ transforms. Recall that solutions $\bar{\psi}$ are obtained from solutions of $H_0 \psi = E \psi$ as in  \eqref{bar psi}, $\psi$ could be either a bound or a scattering state. There is also a BIC $\bar{\psi}_\epsilon$ introduced by the confluent SUSY transformation that also transforms as in \eqref{point phi}; it will be called $\phi_\epsilon(x,t)$. This function solves the time-dependent Schr\"odinger equation $i \partial_t \phi_\epsilon + \partial_{xx}\phi_\epsilon-V\phi_\epsilon=0$ and will be square integrable. Square integrability of $\phi(x,t)$ is guaranteed if the preimage $\bar{\psi}(y)$ is also a square integrable function, 
\begin{eqnarray} \label{square int x}
|| \phi ||^2 = \int_{-\infty}^\infty |\phi(x)|^2 dx = \frac{1}{4t + c_1} \int_{-\infty}^\infty \Bigg| \bar{\psi} \left(\frac{2x + c_2}{2(4t + c_1)}  \right) \Bigg|^2 dx = \int_{-\infty}^\infty \big| \bar{\psi}(y) \big| ^2 dy = || \bar{\psi} ||^2, 
\end{eqnarray}
where we used the change of variable \eqref{yvalue}. %The function $\phi_\epsilon(x,t)$ is then square integrable if $\bar{\psi}_\epsilon(y)$ is also. 
Neither $\phi(x,t)$ nor $\phi_\epsilon(x,t)$ are stationary states, they evolve in time, and they are not eigenfuncions of the operator $-\partial_{xx} + V$. 

Our next step is to select a freezing time $t_F$. At any time $t \geq t_F$, the functions $\phi(x,t_F)$ and $\phi_\epsilon(x,t_F)$ satisfy the eigenvalue equation:  
\begin{eqnarray} \label{Schrodinger magnetic}
\left[ \left( -i\frac{\partial}{\partial x}  +  A_x(x) \right)^2 + V(x,t_F) \right] \phi(x,t_F) = \frac{E}{(4t_F + c_1)^2} \phi(x,t_F), %\quad t \geq t_F,
\end{eqnarray}
where $A_x(x)=- \partial_x g(x)= - (2x-c_2)/(4 t_F+c_1)$ and 
\begin{equation}
    g(x)= \frac{1}{4t_F + c_1} \left(x^2 - c_2 x + \frac{E+c_2^2}{4} \right) 
\end{equation}
is the phase accompanying the wavefunctions \eqref{point phi}.   
Equation \eqref{Schrodinger magnetic} is the Schr\"odinger equation of a charged particle in a magnetic field with vector potential $\mathbf{A}=(A_x,0,0)$, but a null magnetic field since $\mathbf{B}= \nabla \times \mathbf{A}=0$. Here use the gauge transformation introduced in Section \ref{Sec Gauge}, that allows introducing a vector potential $\mathbf{A}(x,t)=(A_x(x,t),0,0)$ where $A_x(x,t)=-\Theta(t-t_F) \partial_x g(x)$, then the piecewise function
 \begin{eqnarray} 
    \phi_F(x,t) =  \begin{cases} 
      \phi(x,t) & 0 \leq t< t_F, \\
      \bar{\psi}\left(\frac{2x + c_2}{2(4t_F+c_1)}\right) \exp \left(- \frac{i E}{(4t_F + c_1)^2} t \right) &  t\geq t_F. 
  \end{cases} \label{phiF}
    \end{eqnarray}
will be solution of 
$$i\partial_t  \phi_F(x,t) =\left[ - \partial_{xx}+V_F(x,t)\right]\phi_F(x,t) =H  \phi_F(x,t).$$ 
In particular, the function 
 \begin{eqnarray}
    \phi_{F\epsilon}(x,t) =  \begin{cases} 
      \phi_\epsilon(x,t) & 0 \leq t< t_F, \\
      \bar{\psi}_\epsilon \left(\frac{2x + c_2}{2(4t_F+c_1)}\right) \exp \left(- \frac{i \epsilon }{(4t_F + c_1)^2} t \right) &  t\geq t_F, 
  \end{cases} \label{Fbic}
    \end{eqnarray}
is a time-dependent wave packet before the freezing time, but after $t_F$, it will become a bound state in the continuum satisfying the eigenvalue equation $H \phi_{F\epsilon}= \varepsilon \phi_{F\epsilon}$, where $\varepsilon = \epsilon/(4 t_F + c_1)^2$.

%%%%%%%%%%%%%%%%%%%%%%%%%
%%%%%%%%%%%%%%%%%%%%%%%%
\section{Examples}\label{Examples}
%%%%%%%%%%%%%%%%%%%%%%%%%%%
%%%%%%%%%%%%%%%%%%%%%%%%%%%
In this section, we construct two potentials with freezable bound states in the continuum. In the first example, we start from the Free Particle defined in the semiaxis and generate a time-dependent potential with a single freezable BIC. For a second example, we show that we can iterate the algorithm to construct a potential with two freezable BICs. Moreover,  using the time-reversal symmetry, we build more potentials with freezable BICS.  

%%%%%%%%%%%%%%%
\subsection{Adding a single freezable bound state in the continuum to the Free-Particle} \label{sec: example 1}
%%%%%%%%%%%%%%

Let us commence our discussion by considering the Free-Particle potential $V_0(y) = 0$ defined in the positive semi-axis $y\in(0,\infty)$. We choose a factorization energy $\epsilon= k^2 > 0$ and $u(y) = \sin(k  y)$. By using a confluent supersymmetric transformation, the potential $V_0$ transform as in \eqref{SUSY C V2}. Explicitly,
\begin{equation}
	V_2(y) = \frac{16  k^2 \left[1-k (2 \omega +y) \sin (2 k y)-\cos (2 k y)\right]}{\left[\sin (2 k y)-2 k
		(2 \omega +y)\right]^2}.
	\label{ussfree}
\end{equation}
To have a regular potential we use $y_0=0$ and $\omega > 0$. The missing state \eqref{bar missing} associated to the factorization energy $\epsilon$ reads
\begin{equation} 
	\bar{\psi}_{\epsilon}(y)=C_\epsilon~ \frac{4 k \sin (k y)}{2 k (2 \omega +y)-\sin (2 k y)},
	\label{psiss3}
\end{equation}
and it is square integrable. To verify this statement, let us focus on the oscillating tail of the function. We can see that the square of the missing state is bounded from above by a square integrable function from $y=\pi/ 4k$ to $y \rightarrow \infty$, as follows:
\begin{align} \label{square int y}
 || \bar{\psi}_\epsilon||^2 &= |C_\epsilon|^2 \left(\int^{ \frac{\pi}{4k}}_0 \left|  \bar{\psi}_\epsilon (y) \right|^2 dy +\int_{ \frac{\pi}{4k}}^\infty \left|  \bar{\psi}_\epsilon (y) \right|^2 dy \right) \nonumber \\
 & = |C_\epsilon|^2  \left( \frac{\pi -2}{\omega (8k \omega + \pi -2)} +\int_{\frac{\pi}{4k}}^\infty  \left|\frac{4 k \sin (k y)}{2 k (2 \omega +y)-\sin (2 k y)}\right|^2 dy  \right) \nonumber \\
 & \leq |C_\epsilon|^2 \left(\frac{\pi -2}{\omega (8k \omega + \pi -2)} +  \int_{\frac{\pi}{4k}}^\infty \left|\frac{4k}{2k(2\omega+y)-1}\right|^2 dy  \right) \nonumber \\
&= |C_\epsilon|^2 \left(\frac{1}{\omega} + \frac{8 k}{8 k \omega + \pi - 2} \right).
\end{align}

\begin{figure}[t]
  \begin{center}
  \includegraphics[width= 7.5 cm ]{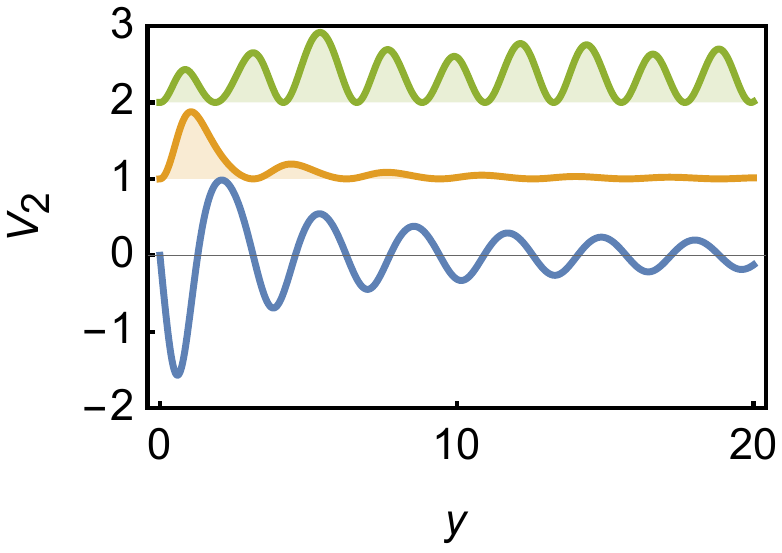}
\caption{Confluent SUSY partner potential $V_2(y)$ of the free particle (blue curve) and probability densities $|\bar\psi_{\epsilon}(y)|^2$ (orange) and $|\bar\psi(y)|^2$ (green) for parameters $\omega=1$, $\epsilon=1$ and $q^2=2$.}\label{Fig1}
   \end{center}
\end{figure}

For an energy  $E_q=q^2 \neq \epsilon $, the  wavefunction $\bar{\psi}(y)$, see \eqref{bar psi}, is
\begin{equation}
	\bar\psi(y)=\frac{4 k \sin ^2(k y) \left[k \cot (k y) \sin (q y)-q \cos (q
		y)\right]}{\left(q^2-k^2\right) \left[2 k (2 \omega +y)-\sin (2 k y)\right]}-\sin (q
	y).
	\label{psiiss3a}
\end{equation}
In Fig.~\ref{Fig1} the potential $V_2(y)$, along with the probability densities of the missing state $\bar\psi_{\epsilon}(y)$ and a scattering state $\bar\psi(y)$ are shown, for $\omega=1$.  We observe that the wavefunction of the BIC has an envelop function that goes to zero as $y\to\infty$, whereas the state $\bar\psi(y)$ is not localized.

Next, we use the point transformation presented in (\ref{yvalue}--\ref{point V}), where we set $c_1=1,~c_2=0$ and $t \in [0, \infty)$. This selection makes $x=y$ at $t=0$. Then $y=x/(4t +1)$ and $V_2$ transforms as:      
\begin{equation}
V(x,t)=\frac{16 k^2 \left[1-k \left(2 \omega +\frac{x}{4 t+1}\right) \sin\left(\frac{2 k x}{4 t+1}\right)-\cos \left(\frac{2 k x}{4t+1}\right)\right]}{\left[(4 t+1) \sin \left(\frac{2 k x}{4 t+1}\right)-2 k (2
	\omega +8 \omega  t+x)\right]^2}.
\label{ussst3}
\end{equation}
Analogously, for the time-dependent BIC, the associated wavefunction for energy $\epsilon $ is explicitly
\begin{equation}
\phi_{\epsilon}(x,t)=\frac{4i k  \sqrt{4 t+1} \exp\Big[{\frac{i \left(4 x^2+k^2  \right)}{16 t+4}}\Big] \sin\left(\frac{k x}{4 t+1}\right)}{(4 t+1) \sin\left(\frac{2 k x}{4 t+1}\right)-2 k (2 \omega +8 \omega  t+x)}.
\label{psiesst3}
\end{equation}
The state is localized and the first maximum in the probability density broadens and diminishes in height as time increases. For states with energy $E=q^2 \ne \epsilon $, the corresponding time-dependent wavefunction has the explicit form
\begin{equation}
\phi(x,t)=\frac{\exp\Big[{\frac{i \left(4 x^2+q^2\right)}{16 t+4}}\Big]}{i\sqrt{4 t+1}}\left[\chi_q(x,t)-\sin
\left(\frac{q x}{4 t+1}\right)\right],
\label{psiisst3}
\end{equation}
where
\begin{equation*}
\chi_q(x,t)=\frac{4 k \sin ^2\left(\frac{k x}{4 t+1}\right) \left[k \cot \left(\frac{k x}{4t+1}\right) \sin \left(\frac{q x}{4 t+1}\right)-q \cos
	\left(\frac{q x}{4 t+1}\right)\right]}{(q^2-k^2) \left[4 k
	\omega +\frac{2 k x}{4 t+1}-\sin \left(\frac{2 k x}{4 t+1}\right)\right]}.
\end{equation*}
This state is unlocalized at any time.

\begin{figure}[t]
 %\centering
 \begin{center}
    \subfloat[]{
       \includegraphics[width= 5 cm ]{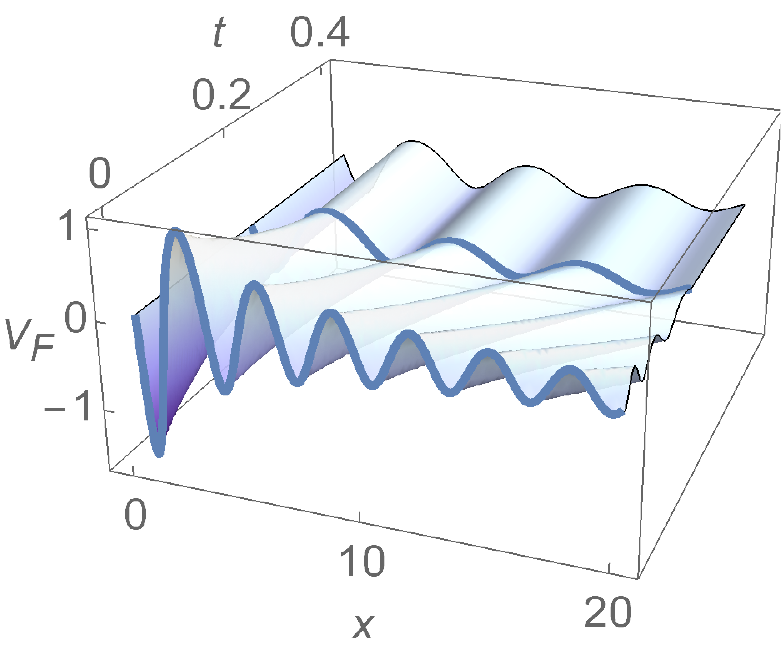}
        \label{Fig2a}} 
  \subfloat[]{
      \includegraphics[width= 5 cm ]{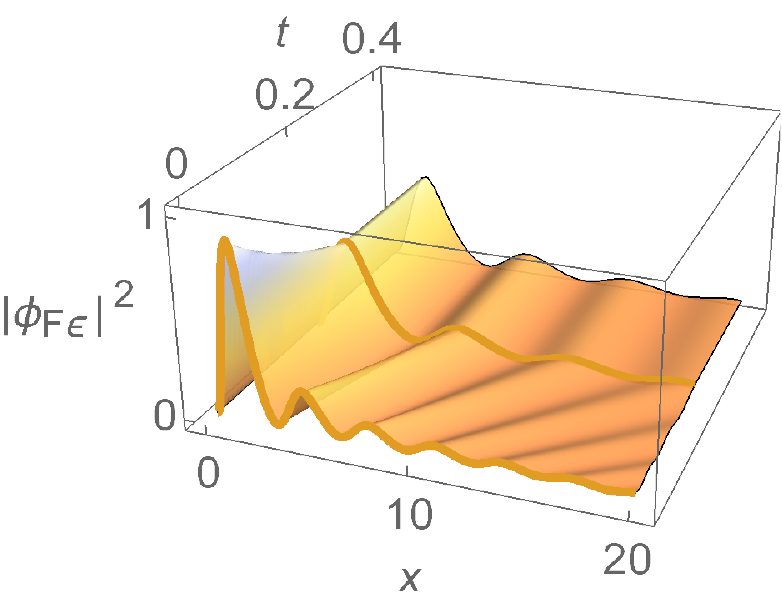}  \label{Fig2b}}
  \subfloat[]{
      \includegraphics[width= 5cm ]{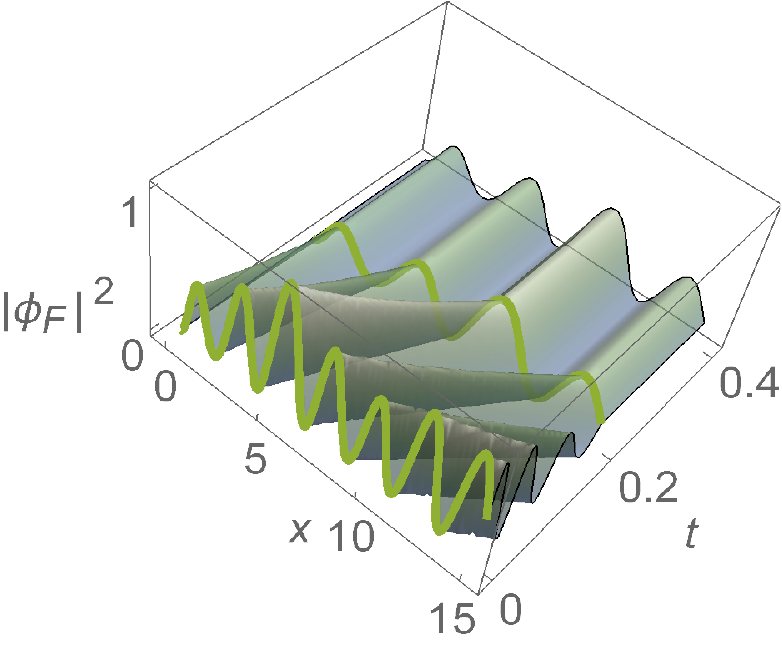}   \label{Fig2c}}
\caption{System with a single freezable BIC. The potential $V_F(x,t)$ (a) and probability densities of $\phi_{F\epsilon} (x,t)$ (b) and $\phi_F(x,t)$ (c). Here $\omega=1$, $\epsilon=1$, $q^2=2$, and $t_F=0.2$. } \label{Fig2}
   \end{center}
 \end{figure}

Finally, we consider a charged particle in the potential:  
 \begin{eqnarray}
    V_F(x,t) =  \begin{cases} 
      V(x,t) & 0 \leq t< t_F, \\
      V(x,t_F) &  t\geq t_F. 
  \end{cases}
    \end{eqnarray}
where $V(x,t)$ is given by \eqref{ussst3}, and $t_F$ is the freezing time. The form of the potential at $t=0$ is oscillatory in the whole domain $x\in (0,\infty)$, the amplitude of such oscillations decrease as $1/x$. This potential is in fact a family parametrized by $\omega > 0$. The smaller the value of $\omega$ the  deeper the first minimum  of the potential.  The solutions of the time dependent Schr\"odinger equation $i\partial_t \phi_F + \partial_{xx} \phi_F + V_F \phi_F=0$ can be constructed as in \eqref{phiF}, \eqref{psiiss3a}  and \eqref{psiisst3}, these states are non-normalizable.  Moreover, the state $\phi_{F \epsilon}$ ends as a bound state in the continuum. It is constructed as in \eqref{Fbic} where $\bar{\psi}_\epsilon$ is given in \eqref{psiss3} and $\phi_\epsilon$ in \eqref{psiesst3}. It is square integrable for all times $t \geq 0$ because of relations \eqref{square int x} and \eqref{square int y}. When $t \geq t_F$, the state $\phi_{F \epsilon}$ becomes the only stationary bound state of the Hamiltonian $H_F= - \partial_{xx} + V_F$ with energy $\epsilon_F= \epsilon/(4t+1)^2$. Since $\epsilon_F > V(x,t_F)$ when $x \to \infty$, then $\phi_{F \epsilon}$ is a freezable bound state in the continuum. In Fig.~\ref{Fig2a}, we show the potential $V_F(x,t)$. Its shape changes in time and its spatial profile oscillates as expected, vanishing as $x\to\infty$. Fig.~\ref{Fig2b} shows the probability density of the added freezable BIC, $|\phi_{F\epsilon}(x,t)|^2$, it can be seen how it varies in time until $t_F=0.2$, when it becomes stationary. The behavior of $|\phi_F(x,t)|^2$, for $E=2$ at different times is shown in Fig.~\ref{Fig2c}. 

%%%%%%%%%%%%
%%%%%%%%%%%%

%%%%%%%%%%%%%%%%%%%%%%%%%%%
\subsection{System with two freezable bound states in the continuum}
%%%%%%%%%%%%%%%%%%%%%%%%%%%%%

We can iterate the confluent SUSY transformation to add more than one bound states in the continuum. With every iteration the length of the expressions of the SUSY partner potential and the solutions of the corresponding Schr\"odinger equation could dramatically increase. To illustrate the procedure, we take the stationary system found in the previous example with the potential term $V_2(y)$ as in \eqref{ussfree}, the single BIC $\bar{\psi}_\epsilon (y)$ with energy $\epsilon=k^2$ as in \eqref{psiss3}, and scattering states $\bar{\psi}(y)$ with energy $E_q=q^2 \neq k^2$ given by   \eqref{psiiss3a}. To simplify notation, let us make the following replacements $\epsilon \rightarrow \epsilon_{1} ,~ k \rightarrow k_1$ and $\omega \rightarrow \omega_{1}$.  Next, we find a confluent-SUSY partner of $V_2$. Since we want add a second BIC, we need to select a second factorization energy $\epsilon_2= k_2^2 \neq \epsilon_1$. The seed solution of this transformation will be the scattering state associated to $\epsilon_2$: 
\begin{eqnarray}
u_2(y)=\frac{L_2^+L_1^+\bar\psi(y) }{k_2^2-k_1^2}  =\frac{4 k_1 \sin ^2(k_1 y) \left[ k_1 \cot (k_1 y) \sin (k_2 y)-k_2 \cos (k_2 y)\right]}{\left(k_2^2-k_1^2\right) \left[2 k_1 (2 \omega_{1} +y)-\sin (2 k_1 y)\right]}-\sin (k_2
	y).
\end{eqnarray}

From \eqref{SUSY C V2}, we can see that the SUSY partner potential of $V_2$ becomes
\begin{eqnarray} \label{SUSY C V4}
V_4(y) =  V_2(y) - 2 \frac{d^2}{dy^2} \left(\omega_{2} +  \int^y_0 u_2^2 dz   \right).
\label{v4}
\end{eqnarray}
The integral in the previous expression can be calculated analytically, unfortunately the explicit expression of $V_4(y)$ is too long to show it in this article. This potential depends on the parameters $\omega_{1}$ and $\omega_{2}$, different values of these parameters gives different potentials, in other words $V_4$ is a biparametric family of SUSY partner potentials of the Free Particle.    

\begin{figure}[t] 
 \begin{center}
      \includegraphics[width= 7.5cm ]{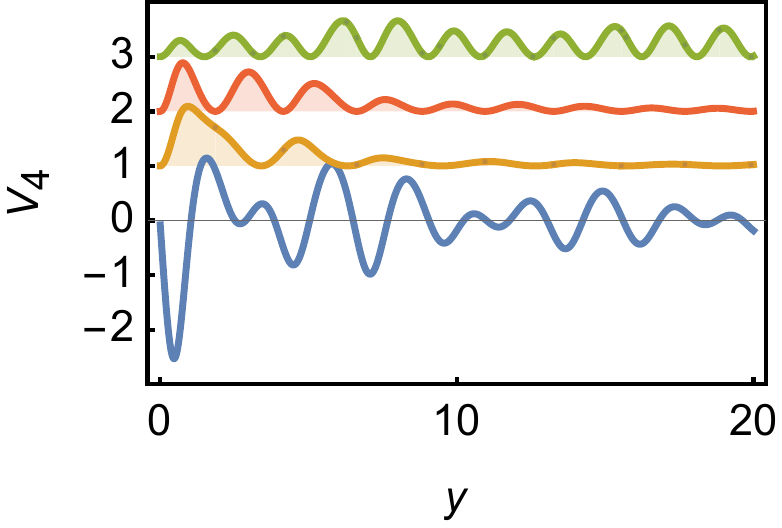}
\caption{Confluent SUSY partner potential $V_4(y)$ of the free particle (blue line) and probability densities $|\widetilde{\psi}_{\epsilon_1}(y)|^2$ (orange), $|\widetilde{\psi}_{\epsilon_2}(y)|^2$ (red),  and $|\widetilde \psi(y)|^2$ (green). The parameters take the values $\omega_1=1,~\omega_2 = 2, ~k_1^2=1,k_2^2=2,~q^2=3,$ and $t_F=0.2$. } \label{Fig3}
  \end{center} 
 \end{figure}

The first BIC correspond to the missing state \eqref{bar missing} takes the form:
\begin{eqnarray}
\widetilde\psi_{\epsilon_2}(y)=\frac{u_2}{\omega_{2}+\int^y_{0} u_2^2 dz}=\frac{1}{\omega_{2}+\int^y_{0} u_2^2 dz}\frac{L_2^\dag L_1^\dag \bar\psi(y)}{( k_2^2-k_1^2)}, 
\label{psie2}
\end{eqnarray}
and satisfies the eigenvalue equation $H_4 \widetilde\psi_{\epsilon_2}= \epsilon_2 \widetilde\psi_{\epsilon_2}$, where $H_4= - \frac{d^2}{dy^2} + V_4$. To construct the second BIC and the scattering states, we need the intertwining operators of this transformation. Analogous to equations \eqref{def H1} and  \eqref{SUSY L}, the operators are \begin{equation}
L_3^\dag=-\frac{d}{dy}+\frac{u'_2}{u_2}, \qquad L_4^\dag =-\frac{d}{dy}+\frac{ v_2'}{ v_2},
\end{equation}
where
\begin{equation}
v_2=\frac{1}{ u_2} \left ( \omega_{2}+\int^y_{0} u_2^2 dz \right). 
\end{equation}
We can obtain the second BIC applying the compose operator $L_4^\dag L_3^\dag$ onto the missing state $\bar \psi_\epsilon (y)$, see  \eqref{psiss3}:
\begin{equation}
    \widetilde\psi_{\epsilon_1}(y)=\frac{1}{(k_1^2-\hat k_2^2)}L_4^\dag L_3^\dag \bar\psi_\epsilon(y),
    \label{psie1}
\end{equation}
it satisfies $H_4 \widetilde\psi_{\epsilon_1}= \epsilon_1 \widetilde\psi_{\epsilon_1}$.  
Finally, the scattering states with energy $E\ne \epsilon_i$, $i=1,2$ are: 
\begin{eqnarray}
\widetilde \psi (y) = \frac{1}{(E-k_2^2)(E-k_1^2)} L_4^\dag L_3^\dag L_2^\dag L_1^\dag \bar \psi (y),
\label{psiefd}
\end{eqnarray}
satisfying the Schr\"odinger equation $H_4 \widetilde \psi = E \widetilde\psi$. 
Lamentably, the explicit expressions of the BICS $\widetilde \psi_{\epsilon_1},~\widetilde \psi_{\epsilon_2}$ and the scattering states $\widetilde \psi$ of the Hamiltonian $H_4(y)$ are too long to write them in this article. In Fig. \ref{Fig3}  we show the plot of the potential $V_4$ (blue curve), the added BICs $\widetilde{\psi}_{\epsilon_1}$ (orange curve), $\widetilde{\psi}_{\epsilon_2}$ (red curve), and a scattering state $\widetilde{\psi}$ (green curve). The parameters we use are  $\omega_1=1,~\omega_2 = 2$, factorization energies $k_1^2=1,k_2^2=2$, and the energy of the scattering state is $q^2=3$.  To show that $\widetilde{\psi}_{\epsilon_i},~i=1,2$ are square integrable functions, we find a square integrable envelope of the form  
\begin{equation}
\left|\frac{a_i}{b_i+y}\right| \geq |\tilde\psi_{\epsilon_i}(y)|.
\end{equation}
Using numerical methods, it is found that $a_1=2.71104$, $b_1=6.79476$, $a_2=2.47686$, and $b_2=8.38096$ give an appropriate fit.

\begin{figure}[t]
 %\centering
 \begin{center}
  \subfloat[]{
   \label{Fig4a} 
    \includegraphics[width= 6cm ]{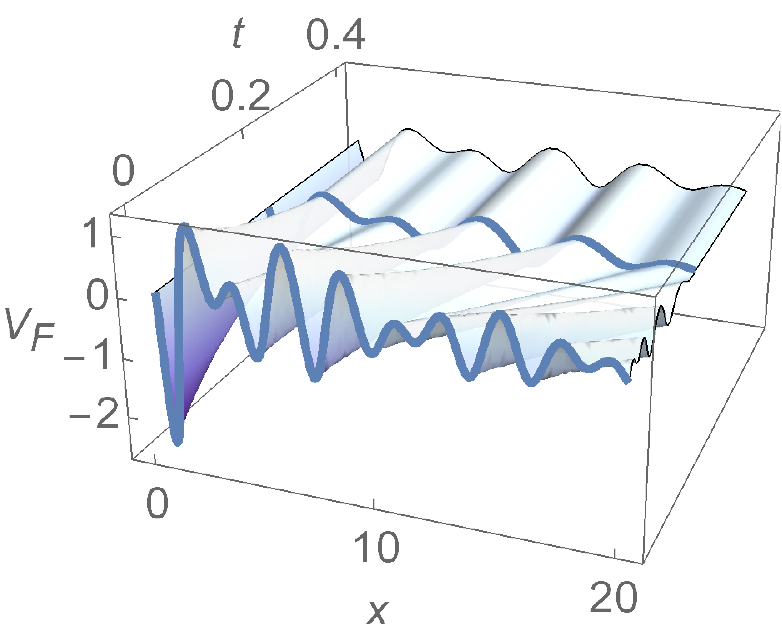}}
     \subfloat[]{
   \label{Fig4b}
      \hspace{1cm} \includegraphics[width= 6.5cm ]{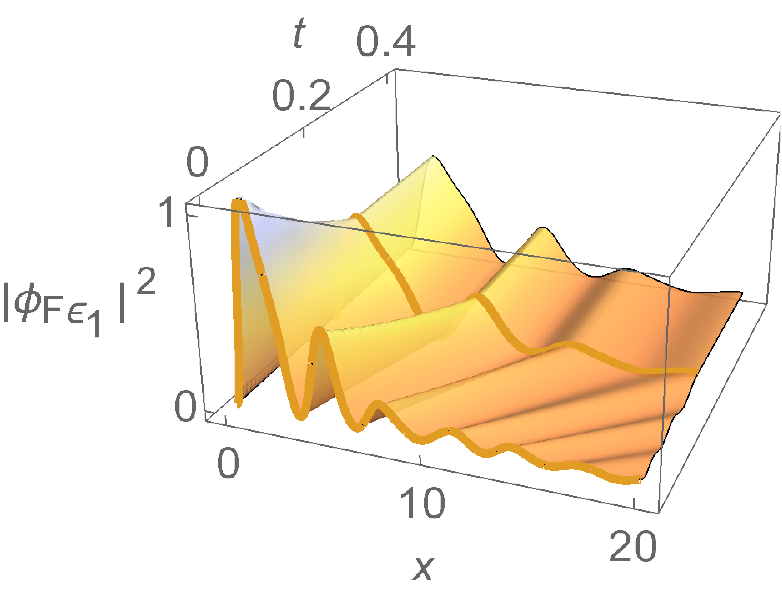}} \\ 
  \subfloat[]{
   \label{Fig4c}
     \includegraphics[width= 6.5cm ]{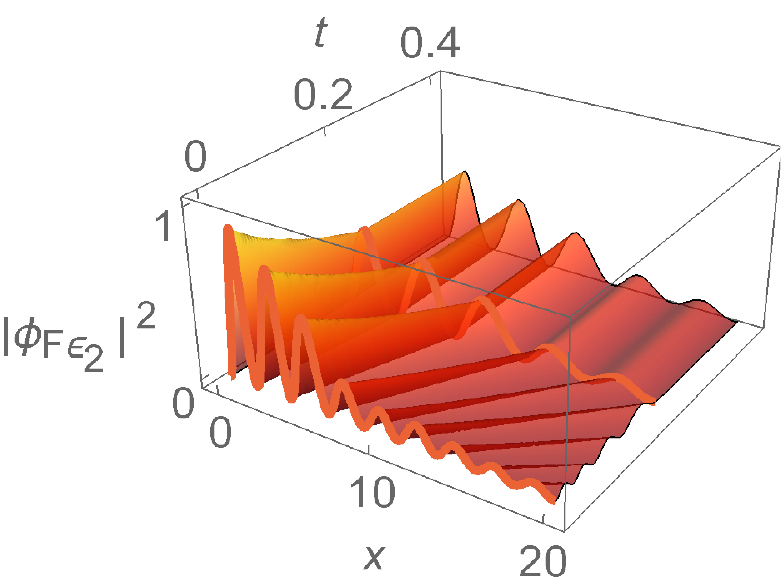}}
    \subfloat[]{ \label{Fig4d}
        \hspace{1cm} \includegraphics[width= 6.5cm ]{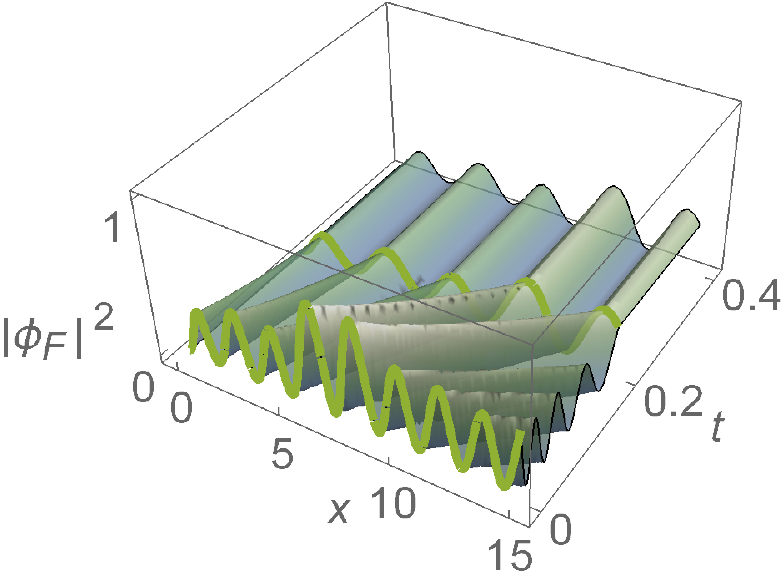}}
\caption{System with two freezable BICs.   Potential $V_F(x,t)$ (a) and probability densities of $\phi_{F\epsilon_1} (x,t)$ (b), $\phi_{F\epsilon_2} (x,t)$ (c), and $\phi_F(x,t)$ (d) at different times $t\geq 0$. The parameters take the values $\omega_1=1,~\omega_2 = 2, ~k_1^2=1,k_2^2=2,~q^2=3,$ and $t_F=0.2$. }
 \label{Fig4}
  \end{center}
 \end{figure}

As in the previous example, we use the point transformation presented in Section \ref{Point T} to obtain a time-dependent potential from  \eqref{v4} and its wavefunctions from (\ref{psie2}, \ref{psie1}, \ref{psiefd}). Recall that we use  $c_1=1,~ c_2=0$, then $t \in [0,\infty)$. The next step is to choose a freezing time $t_F$, then we construct the time-piecewise potential \eqref{Sec 3 VF}. Now, we consider again a vector potential $\mathbf{A}=(A_x,0,0)$ where $A_x(x,t)= -\Theta(t-t_F) 2x / (4 t_F+1)$. The solutions of the time-dependent Schr\"odinger equation associated to $V_F$ are constructed as in  \eqref{phiF}, for scattering states, or  \eqref{Fbic}, for BICS. The energies of the freezable BICs after $t_F$ are $\epsilon_{Fi}= \epsilon_i/ (4 t_F+1),~i=1,2$. Using a freezing time $t_F= 0.2$, we plot $V_F$ in Fig. \ref{Fig4a}, and the probability densities of the freezable BICS $\phi_{F \epsilon_1}$ in Fig. \ref{Fig4b}, $\phi_{F \epsilon_2}$ in Fig. \ref{Fig4c}, and a scattering state $\phi_{F}$ in Fig. \ref{Fig4d}.  

%%%%%%%%%%%%
%%%%%%%%%%%%

\subsubsection{Time-reversal symmetry}

\begin{figure}[t]
 %\centering
 \begin{center}
  \subfloat[]{
      \label{Fig5a}
    \includegraphics[width= 6cm ]{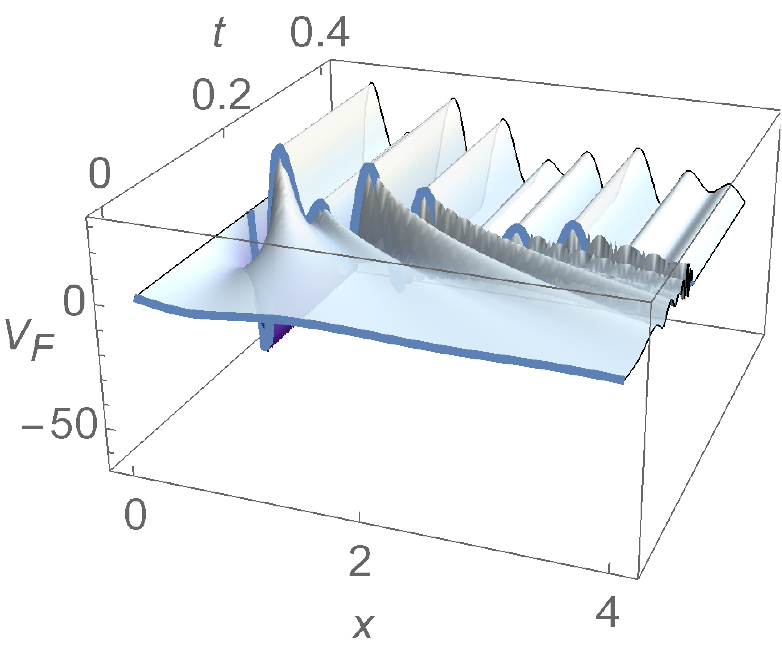}}
    \hspace{10mm}
  \subfloat[]{
   \label{Fig5b}
    \includegraphics[width= 6.5cm ]{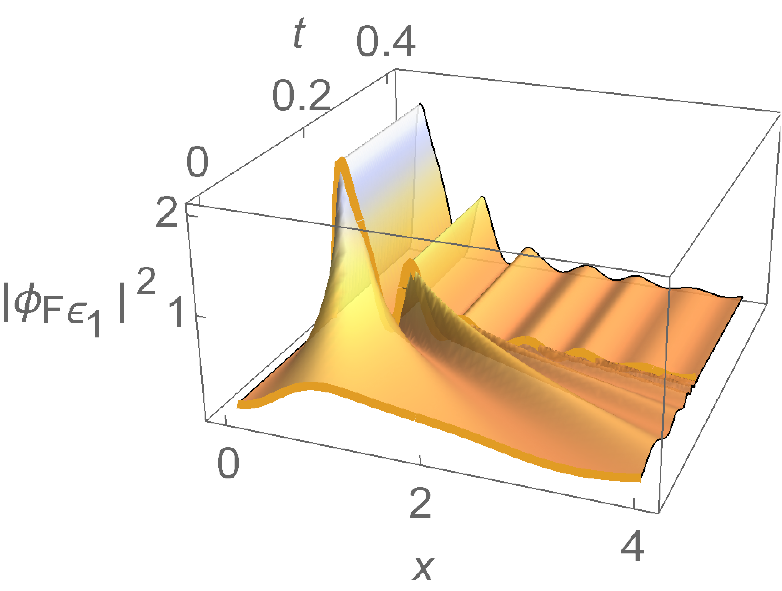}} \\
  \subfloat[]{
   \label{Fig5c}
     \includegraphics[width= 6.5cm ]{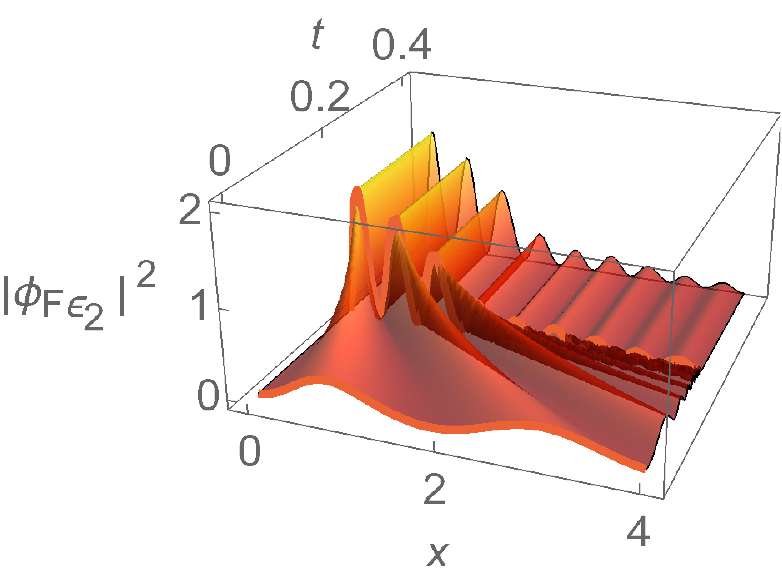}}
     \hspace{10mm} 
     \subfloat[]{
     \label{Fig5d}
     \includegraphics[width= 6.5cm ]{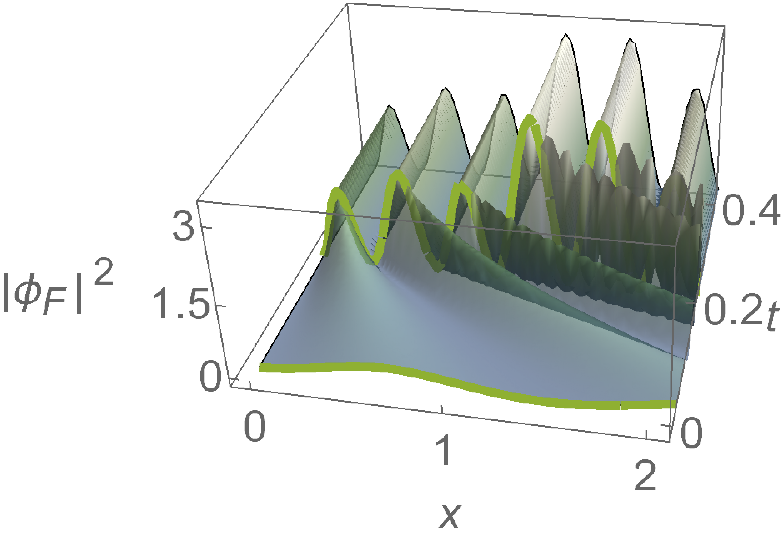}}
\caption{System with two BICs using the temporal inversion symmetry. Potential $V_F(x,t)$ (a), and probability densities of $\phi_{F\epsilon_1} (x,t)$ (b), $\phi_{F\epsilon_2} (x,t)$ (c) and $\phi_F(x,t)$ (d). The parameters take the values  $\omega_1=1,~\omega_2 = 2, ~k_1=1,k_2^2=2,~q^2=3$ and $t_F=0.2$. } \label{Fig5}
 \label{f:dynamic}
  \end{center}
 \end{figure}

Since our system is Hermitian, there is a time-reversal symmetry. By taking the complex conjugate of the Schr\"odinger equation, and replacing $t \to - t$ we can see that $\phi_F^*(x,-t)$ solves the time-dependent Schr\"odinger equation for the potential $V_F(x,-t)$. The advantage of using this transformation is that at $t=0$ the amplitude of the oscillations are smaller than at $t=t_F$, in fact, if choosing correctly the parameter $c_1$ (recall that in this example we used $c_1=1$), the shape of the potential $V_F(x,0)$ resembles the free-particle potential and the frozen potential will present oscillations with greater amplitudes. In Fig. \ref{Fig5} we applied the time-reversal transformation of the example consider to make Fig. \ref{Fig4}. It can be seen that the potential $V_F$ in Fig. \ref{Fig5a} has a flat shape at $t=0$ and then the oscillations become more visible and narrower as $t$ increases,  the same is true for the freezable BICs $\phi_{F\epsilon_1}$ in Fig. \ref{Fig5b}, $\phi_{F\epsilon_2}$ in Fig. \ref{Fig5c}, and the scattering state $\phi_{F}$ in Fig. \ref{Fig5d} after the transformation.

%%%%%%%%%%%%
%%%%%%%%%%%%
%%%%%%%%%%%%%%%
%%%%%%%%%%%%%%%

%%%%%%%%%%%%%%%%%%%%%%%
\section{Final remarks}\label{FR}
%%%%%%%%%%%%%%%%%%%%%%%

In this article, we have made use of supersymmetric confluent transformations. Starting from a stationary system without bound states in the continuum, we have generated stationary potentials that support a localized, squared integrable state, the BIC at certain factorization energy embedded in the continuum spectrum. For any other energy value in the continuum, the corresponding state is extended and corresponds to a scattering state. Through a point transformation, we have provided the potential and states with time evolution.  Nevertheless, we notice that the wrinkles in the potential as $x\to\infty$ still localize a BIC at every fixed time. 

Next, we allow the evolution of the system to continue, and at a given time, we freeze the potential such that it no longer evolves but remains stationary. We then study the behavior of the BIC with this static potential after the freeze-out time. We notice that this state is not a solution of the stationary Schr\"odinger equation, but instead, it develops a geometric phase in terms of a vector potential that does not generate any magnetic field whatsoever. This observation allows us to gauge out this geometric phase and thus observe that the resulting state becomes indeed is an eigenstate of the frozen Hamiltonian corresponding to a BIC.

We further show that the presented procedure can be iterated to add extra BICs at different factorization energy. Expressions can be lengthy and cumbersome, though straightforward to derive. The stationary multiple-BIC system can be granted a time evolution via point transformations up to a new freeze-out time where the potential is required to remain stationary. By gauging away the geometric phase developed by the states during the time evolution, we still find the BICs to remain localized by their reflection of the Bragg mirror of the potential.

We show the use of the technique in two examples. We first added a single freezable BIC to the Free Particle defined in the semiaxis. Explicit expressions of the time-dependent potential, scattering states, and the freezable BIC are given. Then, we inserted a second freezable BIC at different energy through an iteration of the confluent supersymmetric transformation. We verify that the family of time-dependent potentials with freezable BICs  can increase using a time-reversal symmetry. Further examples related to spet-like potentials have been explored in~\cite{proceedings}.

A natural extension of these ideas is to consider a relativistic system starting from a Dirac equation. Although quantum BICs still await a true observation, the new class of modern materials might offer a chance to explore these states. All these ideas are under consideration, and results shall be presented elsewhere.

\section*{Acknowledgments}
The authors acknowledge Consejo Nacional de Ciencia y Tecnolog\'{\i}a (CONACyT-M\'exico) under grant FORDECYT-PRONACES/61533/2020.

%\bibliographystyle{unsrt}
%\bibliography{references}

%\section*{Appendix}

\end{document}